# Modeling Closed-loop Analog Matrix Computing Circuits with Interconnect Resistance

Mu Zhou[†], Junbin Long[†], Yubiao Luo, and Zhong Sun, *Senior Member, IEEE*

*Abstract*—Analog matrix computing (AMC) circuits based on resistive random-access memory (RRAM) have shown strong potential for accelerating matrix operations. However, as matrix size grows, interconnect resistance increasingly degrades computational accuracy and limits circuit scalability. Modeling and evaluating these effects are therefore critical for developing effective mitigation strategies. Traditional SPICE (Simulation Program with Integrated Circuit Emphasis) simulators, which rely on modified nodal analysis, become prohibitively slow for large-scale AMC circuits due to the quadratic growth of nodes and feedback connections. In this work, we model AMC circuits with interconnect resistance for two key operations—matrix inversion (INV) and eigenvector computation (EGV), and propose fast solving algorithms tailored for each case. The algorithms exploit the sparsity of the Jacobian matrix, enabling rapid and accurate solutions. Compared to SPICE, they achieve several orders of magnitude acceleration while maintaining high accuracy. We further extend the approach to open-loop matrix–vector multiplication (MVM) circuits, demonstrating similar efficiency gains. Finally, leveraging these fast solvers, we develop a bias-based compensation strategy that reduces interconnect-induced errors by over 50% for INV and 70% for EGV circuits. It also reveals the scaling behavior of the optimal bias with respect to matrix size and interconnect resistance.

## I. INTRODUCTION

Matrix computing is fundamental to a wide range of data-intensive applications. Analog matrix computing (AMC) circuits based on resistive random-access memory (RRAM) provide high parallelism, enabling the rapid execution of matrix operations. With the cross-point array, matrix–vector multiplication (MVM) can be conveniently performed in the open-loop manner [1][2]. By incorporating feedback loops, AMC circuits become closed-loop systems capable of executing a wide range of fundamental matrix operations, such as matrix inversion (INV) and eigenvector computation (EGV) [3]. Unlike digital circuits, which rely on extensive and complex logic designs, AMC circuits perform matrix computations with an $O(1)$ time complexity [4][5][6]. Furthermore, the computation is performed in-situ within the memory array, endowing these circuits with inherent in-memory computing capabilities. Consequently, they are regarded as an effective approach to overcome the von Neumann bottleneck faced by traditional computing architectures [7].

Although cross-point arrays have proven effective in accelerating matrix operations, the inherently high error rates in analog computing continue to pose a significant barrier to practical applications. Compared to other non-ideal factors such as analog noise and device variations, errors caused by interconnect resistance are considerably more difficult to model and evaluate, as this resistance significantly complicates the circuit topology. Besides, as the matrix size increases, the impact of interconnect resistance on computational accuracy becomes more significant and harder to evaluate. Existing SPICE (Simulation Program with Integrated Circuit Emphasis) simulators employ modified nodal analysis (MNA) to analyze circuits [8]. For an $N \times N$ RRAM cross-point array, the inclusion of interconnect resistance increases the number of nodes that SPICE needs to solve from $2N$ to approximately $2N^2$. This increase substantially elevates the time and space overhead for simulation. Specifically, the time complexity increases from approximately $O(N^3)$ to around $O(N^6)$ [9][10], making large-scale simulations computationally prohibitive.

While practical models and iterative algorithms have been proposed for estimating interconnect resistance in MVM circuits [9]-[15], the presence of feedback loops and operational amplifiers (OAs) makes analyzing INV and EGV circuits far more challenging [16]. To date, little research has addressed circuit-level modeling of closed-loop AMC circuits with interconnect resistance, and the absence of efficient simulators hinders efforts to evaluate and mitigate these effects, including the development of compensation strategies for interconnect-induced errors.

In this work, we derive the nodal equations for closed-loop INV and EGV circuits with interconnect resistance and propose generalized solving algorithms that rapidly compute their solutions. These algorithms offer low time complexity while maintaining accuracy comparable to SPICE, achieving substantial speedup in simulations across numerous random matrix equations. Building on this analysis, we further develop a bias-based strategy that applies targeted input adjustments to compensate for interconnect resistance, significantly reducing the circuit's relative error.

This work was supported by Beijing Natural Science Foundation (4252016) and the 111 Project (B18001). *(Corresponding author: Zhong Sun)*
M. Zhou is with School of Electronics Engineering and Computer Science, Peking University, Beijing 100871, China.
J. Long, Y. Luo, and Z. Sun are with Institute for Artificial Intelligence, and School of Integrated Circuits, Peking University, Beijing 100871, China (email: zhong.sun@pku.edu.cn).
Z. Sun is with Beijing Advanced Innovation Center for Integrated Circuits.
[†]These authors contributed equally to this work.

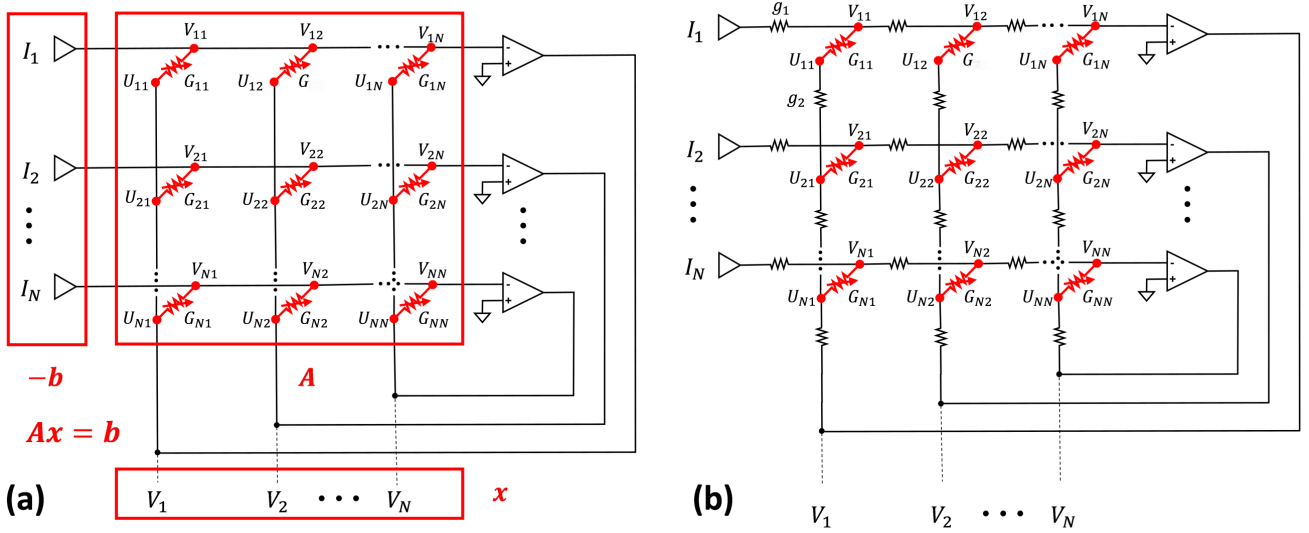

Fig. 1. Schematic of the INV circuit. (a) INV circuit without interconnect resistance. In this ideal case, the circuit solves the matrix equation $Ax=b$. (b) INV circuit with interconnect resistance. The crosspoint array maps a matrix $G$ with conductance values. The input currents represent a column vector $i = [I_1, I_2, …, I_N]$ and the output voltages represent a row vector $v = [V_1, V_2, …, V_N]$.

## II. Modeling INV and EGV Circuits with Interconnect Resistances

### A. INV Circuit Analysis

The fundamental INV circuit, shown in Fig. 1a, is designed to solve the matrix equation:

$$Ax = b \quad (1)$$

where $A$ is an $N \times N$ matrix mapped to the conductance values of the RRAM array, and $b$ is an $N \times 1$ vector mapped to the input currents. Once the circuit reaches steady state, the solution vector $x$ ($N \times 1$) can be obtained from the feedback path.

The INV circuit incorporating interconnect resistance is shown in Fig. 1b. In this circuit, $G_{ij}$ represents the conductance corresponding to the original matrix element, while $r_1 = 1/g_1$ and $r_2 = 1/g_2$ denote the row and column interconnect resistance between two neighboring cells, respectively. The resulting network contains more than $2N^2$ nodes. Solving such a large-scale system with SPICE simulators entails a computational complexity of $O(N^6)$.

To model the INV circuit with interconnect resistance, we assume that the interconnect resistance is uniform across all rows and columns, determined by the device fabrication process [17], and that the OA is ideal with principles of "virtual short" and "virtual open" being valid [18]. In addition, due to the circuit stability limit, the matrix in consideration should be positive definite [4][5]. These assumptions would be validated in the subsequent circuit simulations.

As shown in Fig. 1b, we denote the common node between the RRAM and the row interconnect resistance as $V_{ij}$ ($i,j = 1,2,…,N$), and the common node between the RRAM and the column interconnect resistance as $U_{ij}$. The output voltages of the OAs (i.e., the solution vector $x$) are denoted as $V_i$.

According to Kirchhoff's current law (KCL), the algebraic sum of currents flowing into node $V_{ij}$ is zero. Thus, for column number $j$ greater than 1, the following equation is formulated:

$$G_{ij}(U_{ij} - V_{ij}) = g_1(V_{ij} - V_{i,j-1}) + g_1(V_{ij} - V_{i,j+1}),$$
$$j < N \quad (2)$$
$$G_{ij}(U_{ij} - V_{ij}) = g_1(V_{ij} - V_{i,j-1}) + g_1(V_{ij} - 0), j = N$$

For the first column, due to the presence of the input current source, the equation becomes:

$$G_{ij}(U_{ij} - V_{ij}) = g_1(V_{ij} - V_{i,j-1}) - I_j \quad (3)$$

Let $I_0 = \begin{bmatrix} I_1 & 0 & \cdots & 0 \\ I_2 & 0 & \cdots & 0 \\ \vdots & \vdots & \ddots & \vdots \\ I_N & 0 & \cdots & 0 \end{bmatrix}$, and $D = \begin{bmatrix} 1 & -1 & 0 & \cdots & 0 & 0 \\ -1 & 2 & -1 & \cdots & 0 & 0 \\ 0 & -1 & 2 & \cdots & 0 & 0 \\ \vdots & \vdots & \vdots & \ddots & \vdots & \vdots \\ 0 & 0 & 0 & \cdots & 2 & -1 \\ 0 & 0 & 0 & \cdots & -1 & 2 \end{bmatrix}$. Eq. (2) and Eq. (3) are combined as:

$$G \circ (U - V) = g_1 VD - I_0 \quad (4)$$

where $G$, $U$ and $V$ are $N \times N$ square matrices composed of $G_{ij}$, $U_{ij}$ and $V_{ij}$, respectively, and "$\circ$" denotes the Hadamard product. Similarly, by applying KCL to node $U_{ij}$ and rearranging the terms, we obtain

$$G \circ (V - U) = g_2(DU - V_0) \quad (5)$$

where $V_0$ is defined as $V_0 = \begin{bmatrix} 0 & 0 & \cdots & 0 \\ \vdots & \vdots & \ddots & \vdots \\ 0 & 0 & \cdots & 0 \\ V_1 & V_2 & \cdots & V_N \end{bmatrix}$. Consider the virtual short principle of the OAs, which implies $V_{iN} = 0$, $V$

**Algorithm 1:** INV Algorithm for Cross-point RRAM Array with Interconnect Resistances

**Data:**
Matrix size $N \times N$, where $N$ is the dimension of the square matrix;
Conductance matrix $\boldsymbol{G} \in \mathbb{R}^{N \times N}$;
Row interconnect conductance $g_1$;
Column interconnect conductance $g_2$;
Input current vector $\boldsymbol{i} \in \mathbb{R}^N$.

**Result:**
Output voltage vector $\boldsymbol{v}$.

1 **Step 1:** Generate matrices $\boldsymbol{D}, \boldsymbol{M}_1, \boldsymbol{M}_2$;
2 **Step 2:** Construct matrix $\boldsymbol{I}_0$ based on $\boldsymbol{i}$ and initialize $\boldsymbol{\theta}$;
3 **Step 3:** Calculate the residual $\boldsymbol{\Delta F}$ of function $\boldsymbol{F}$;
4 **Step 4:** Calculate the Jacobian matrix $\boldsymbol{J}$ and solve $\boldsymbol{J}\Delta\boldsymbol{\theta} = -\Delta\boldsymbol{F}$;
5 **Step 5:** Update: $\boldsymbol{\theta} \leftarrow \boldsymbol{\theta} + \Delta\boldsymbol{\theta}$;
6 **Step 6:** Output: $\boldsymbol{v} = (\boldsymbol{\theta} \cdot \boldsymbol{M}_2)^\mathrm{T}[N,:]$;

can be rewritten as: $\boldsymbol{V} = \begin{bmatrix} V_{11} & V_{12} & \cdots & V_{1,N-1} & 0 \\ V_{21} & V_{22} & \cdots & V_{2,N-1} & 0 \\ \vdots & \vdots & \ddots & \vdots & \vdots \\ V_{N1} & V_{N2} & \cdots & V_{N,N-1} & 0 \end{bmatrix}$. By combining Eq. (4) and Eq. (5) and eliminating $\boldsymbol{U}$, we obtain the resulting node equations:

$$\frac{1}{g_2}\boldsymbol{I}_0 + \boldsymbol{V}_0 - \frac{g_1}{g_2}\boldsymbol{V}\boldsymbol{D} - \boldsymbol{D}\boldsymbol{V} - \boldsymbol{D}\left[\frac{1}{\boldsymbol{G}} \circ (g_1\boldsymbol{V}\boldsymbol{D} - \boldsymbol{I}_0)\right] = \boldsymbol{0} \quad (6)$$

where $\boldsymbol{V}_0$ and $\boldsymbol{V}$ are unknown matrices and all other terms are known. The unknown parts of $\boldsymbol{V}$ and $\boldsymbol{V}_0$ can be precisely combined to form an $N \times N$ matrix, which indicates that Eq. (6) is solvable. For further simplification, we define the solution matrix as

$$\boldsymbol{\theta} = \begin{bmatrix} V_{11} & V_{12} & \cdots & V_{1,N-1} & V_1 \\ V_{21} & V_{22} & \cdots & V_{2,N-1} & V_2 \\ \vdots & \vdots & \ddots & \vdots & \vdots \\ V_{N1} & V_{N2} & \cdots & V_{N,N-1} & V_N \end{bmatrix}.$$

By substituting $\boldsymbol{V} = \boldsymbol{\theta}\boldsymbol{M}_1$ and $\boldsymbol{V}_0 = (\boldsymbol{\theta}\boldsymbol{M}_2)^\mathrm{T}$ into Eq. (6), we obtain an equation solely in terms of the unknown $\boldsymbol{\theta}$:

$$\frac{1}{g_2}\boldsymbol{I}_0 + (\boldsymbol{\theta}\boldsymbol{M}_2)^\mathrm{T} - \frac{g_1}{g_2}\boldsymbol{\theta}\boldsymbol{M}_1\boldsymbol{D} - \boldsymbol{D}\boldsymbol{\theta}\boldsymbol{M}_1 \\ -\boldsymbol{D}\left[\frac{1}{\boldsymbol{G}} \circ (g_1\boldsymbol{\theta}\boldsymbol{M}_1\boldsymbol{D} - \boldsymbol{I}_0)\right] = \boldsymbol{0} \quad (7)$$

where $\boldsymbol{M}_1 = \begin{bmatrix} 1 & 0 & \cdots & 0 & 0 \\ 0 & 1 & \cdots & 0 & 0 \\ \vdots & \vdots & \ddots & \vdots & \vdots \\ 0 & 0 & \cdots & 1 & 0 \\ 0 & 0 & \cdots & 0 & 0 \end{bmatrix}$, $\boldsymbol{M}_2 = \begin{bmatrix} 0 & 0 & \cdots & 0 & 0 \\ 0 & 0 & \cdots & 0 & 0 \\ \vdots & \vdots & \ddots & \vdots & \vdots \\ 0 & 0 & \cdots & 0 & 0 \\ 0 & 0 & \cdots & 0 & 1 \end{bmatrix}$.

The solution matrix $\boldsymbol{\theta}$ can be obtained from Eq. (7), donated as $\boldsymbol{F}(\boldsymbol{\theta}) = \boldsymbol{0}$. Since this is a fully linear equation, its Jacobian matrix is constant, allowing us to directly express the solution as

$$\boldsymbol{\theta} = reshape\left[-\boldsymbol{J}_{inv}^{-1}\bigl(vec(\boldsymbol{\theta})\bigr) \cdot \boldsymbol{F}\bigl(vec(\boldsymbol{\theta})\bigr)\right] \quad (8)$$

where $\boldsymbol{J}_{inv}$ is the Jacobian matrix of Eq. (7) with respect to $\boldsymbol{\theta}$, $vec(\boldsymbol{\theta})$ is the row-vectorized form of $\boldsymbol{\theta}$ (a vector of size $N^2 \times 1$), and we use the function $reshape$ to reorganize the results into an $N \times N$ matrix. Based on Eq. (7), we can derive an explicit expression for $\boldsymbol{J}_{inv}$:

$$\boldsymbol{J}_{inv} = (\boldsymbol{I}_N \otimes \boldsymbol{M}_2)^\mathrm{T} - \frac{g_1}{g_2}(\boldsymbol{M}_1\boldsymbol{D} \otimes \boldsymbol{I}_N) - \boldsymbol{M}_1 \otimes \boldsymbol{D} \\ -(\boldsymbol{I}_N \otimes \boldsymbol{D})diag\left(\frac{g_1}{\boldsymbol{G}}\right)(\boldsymbol{M}_1\boldsymbol{D} \otimes \boldsymbol{I}_N) \quad (9)$$

where the operator $diag\left(\frac{g_1}{\boldsymbol{G}}\right)$ denotes a diagonal matrix formed from the element-wise division of the constant $g_1$ by the $N \times N$ matrix $\boldsymbol{G}$. Specifically, $\boldsymbol{G}$ is first flattened into a vector of length $N^2$ in row-major order, denoted as $vec(\boldsymbol{G})$. Then, each element of $\frac{g_1}{vec(\boldsymbol{G})}$ is placed on the main diagonal of an $N^2 \times N^2$ diagonal matrix, with all off-diagonal elements set to zero. Then we have the solution for $\boldsymbol{\theta}$ using Eq. (8) and recover $\boldsymbol{V}_0$ via $\boldsymbol{V}_0 = (\boldsymbol{\theta}\boldsymbol{M}_2)^\mathrm{T}$, thereby obtaining the result of the INV circuit. A summary of the procedure is presented in Algorithm 1. As will be analyzed later in Section III.B, the sparsity of the Jacobian matrix would accelerate the solution of Eq. (8), and thus should Algorithm 1 have an inherently low time complexity.

*B. EGV Circuit Analysis*

The EGV circuit is another important AMC circuit, whose basic structure is shown in Fig. 2a [5]. It is designed to solve the following matrix equation:

$$\boldsymbol{A}\boldsymbol{x} = \lambda\boldsymbol{x} \quad (10)$$

where $\boldsymbol{A}$ is an $N \times N$ matrix mapped to the conductance values of the RRAM array, $\lambda$ is a constant mapped to the conductance value $G_\lambda$, and $\boldsymbol{x}$ is the $N \times 1$ eigenvector to be solved. It can be proven that when $\lambda$ approaches the dominant eigenvalue of matrix $\boldsymbol{A}$, the circuit converges to a stable solution. Therefore, the eigenvalues of matrix $\boldsymbol{A}$ can be determined by sweeping $G_\lambda$ [20]. Once the circuit stabilizes, the eigenvector corresponding to $\lambda$ can be obtained from the feedback path, with additional normalization required.

The EGV circuit model considering interconnect resistance is shown in Fig. 2b. Here, for formal consistency with the INV circuit, we employ an optimized EGV circuit with one of its global feedback paths disconnected and an external voltage input $V_0$ applied. Its effectiveness in solving the eigenvector task is validated through simulation. Using an analytical approach similar to that in Section II.A, the following equation is formulated:

$$g_1\boldsymbol{D}\left(\frac{1}{\boldsymbol{G}} \circ \boldsymbol{V}\boldsymbol{D}\right) + \frac{g_1}{g_2}\boldsymbol{V}\boldsymbol{D} + \boldsymbol{D}\boldsymbol{V} \\ -\frac{g_1}{G_\lambda}(\boldsymbol{V}\boldsymbol{M}_5)^\mathrm{T}\boldsymbol{M}_4 - \boldsymbol{V}_0\boldsymbol{M}_3 = \boldsymbol{0} \quad (11)$$

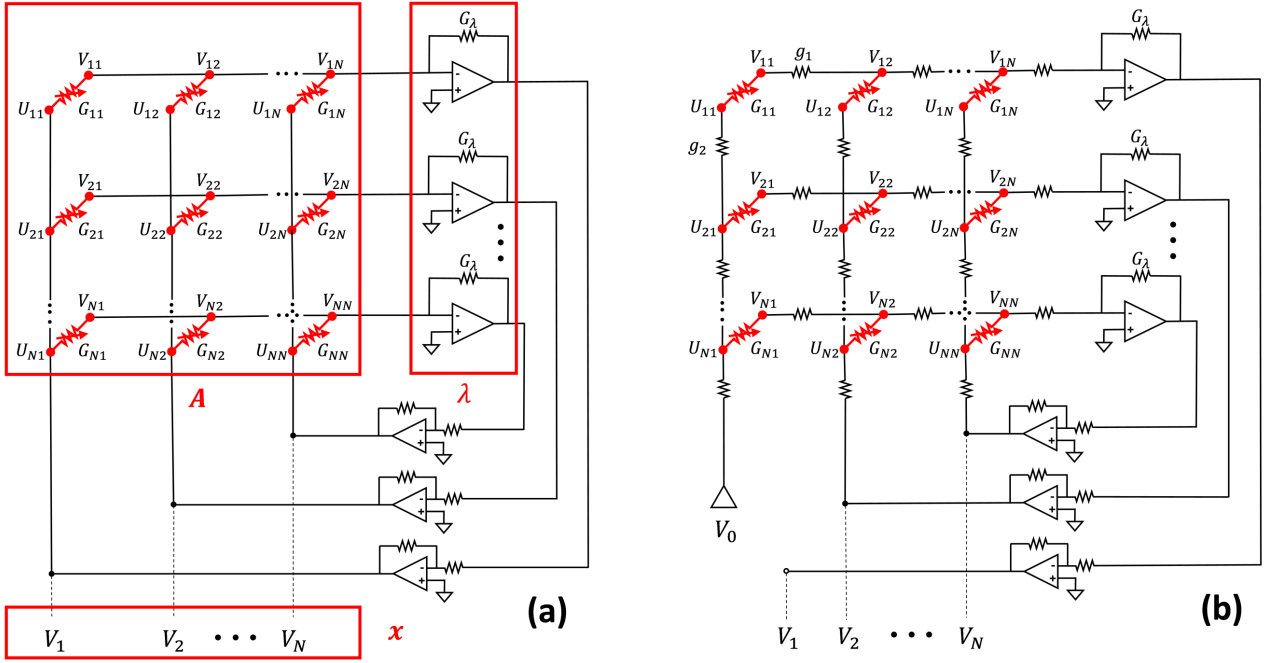

Fig. 2. Schematic of the EGV circuit. (a) EGV circuit without interconnect resistance. In this ideal case, the circuit solves the matrix equation $Ax=\lambda x$. (b) Optimized EGV circuit with interconnect resistance. The crosspoint array maps a matrix $G$ with conductance values and the feedback conductance maps the eigenvalue of matrix $A$. The input voltage represents a small value $V_0$ and the output voltages represent a row vector $v = [V_1, V_2, …, V_N]$.

where $M_3 = \begin{bmatrix} 0 & 0 & \cdots & 0 & 0 \\ 0 & 0 & \cdots & 0 & 0 \\ \vdots & \vdots & \ddots & \vdots & \vdots \\ 0 & 0 & \cdots & 0 & 0 \\ 1 & 0 & \cdots & 0 & 0 \end{bmatrix}$, $M_4 = \begin{bmatrix} 0 & 0 & \cdots & 0 & 0 \\ 0 & 1 & \cdots & 0 & 0 \\ \vdots & \vdots & \ddots & \vdots & \vdots \\ 0 & 0 & \cdots & 1 & 0 \\ 0 & 0 & \cdots & 0 & 1 \end{bmatrix}$,

$M_5 = \begin{bmatrix} 0 & 0 & \cdots & 0 & 0 \\ 0 & 0 & \cdots & 0 & 0 \\ \vdots & \vdots & \ddots & \vdots & \vdots \\ 0 & 0 & \cdots & 0 & 0 \\ 0 & 0 & \cdots & 0 & 1 \end{bmatrix}$, and $D$ is identical to that in Section II.A. $V$ is the only unknown matrix in Eq. (11), which is the potentials of nodes between the RRAM and the row interconnect resistance. For Eq. (11), its Jacobian matrix can be explicitly obtained, namely

$$J_{egv} = (I_N \otimes D) diag\left(\frac{g_1}{G}\right)(D \otimes I_N)$$
$$+ \frac{g_1}{g_2}(D \otimes I_N) + (I_N \otimes D) - \frac{g_1}{G_\lambda}(M_4 \otimes M_5) \quad (12)$$

Then, the solution of Eq. (11) can be expressed as

$$V = reshape[-J_{egv}^{-1}(vec(V)) \cdot F(vec(V))] \quad (13)$$

During the equation solving process, we deliberately enhanced the sparsity of the Jacobian matrix by rearranging terms and avoiding the presence of dense matrices, such as $D^{-1}$. This is crucial for reducing the algorithm's time complexity. Once Eq. (11) is solved, we can obtain the solution eigenvector by

$$V_x = \frac{g_1}{G_\lambda}(VM_3)^T \quad (14)$$

A summary of the above procedure is presented in Algorithm 2. For the same reason as Algorithm 1, Algorithm 2 would also achieve a speedup compared to iterative algorithms.

**Algorithm 2:** EGV Algorithm for Crosspoint RRAM Array with Interconnect Resistances

**Data:**
Matrix size $N \times N$, where $N$ is the dimension of the square matrix;
Conductance matrix $G \in \mathbb{R}^{N \times N}$;
Row interconnect conductance $g_1$;
Column interconnect conductance $g_2$;
Feedback conductance $G_\lambda$ corresponding to eigenvalue $\lambda$.

**Result:**
Output voltage vector $v$.

1 **Step 1:** Generate matrices $D$, $M_3$, $M_4$, $M_5$;
2 **Step 2:** Initialize $V$;
3 **Step 3:** Calculate the residual $\Delta F$ of function $F$;
4 **Step 4:** Calculate the Jacobian matrix $J$ and solve $J\Delta V = -\Delta F$;
5 **Step 5:** Update: $V \leftarrow V + \Delta V$;
6 **Step 6:** Output: $v = \left(\frac{g_1}{G_\lambda}\right) \cdot (V \cdot M_5)^T [N,:]$;

*C. Extension to MVM Circuits*

Although the algorithms mentioned above are developed for closed-loop AMC circuits, a similar algorithm can also be applied to the open-loop MVM circuit. Prior work in [9]-[15] have derived the matrix equations for MVM circuits with interconnect resistance, and the work in [15] employed a fixed-point iteration algorithm for their solution. We derive the Jacobian matrix of the final matrix equation presented in

**Algorithm 3:** MVM Algorithm for Cross-point RRAM Array with Interconnect Resistances

**Data:**
Matrix size $M \times N$, where $M$ and $N$ denote the number of rows and columns, respectively;
Conductance matrix $\boldsymbol{G} \in \mathbb{R}^{N \times N}$;
Row interconnect conductance $g_1$;
Column interconnect conductance $g_2$;
Input voltage vector $\boldsymbol{v} \in \mathbb{R}^N$.

**Result:**
Output current vector $\boldsymbol{i}$.

1. **Step 1:** Generate matrices $\boldsymbol{D}$, $\boldsymbol{D}_1$;
2. **Step 2:** Initialize $\boldsymbol{U}$;
3. **Step 3:** Calculate the residual $\Delta \boldsymbol{F}$ of function $\boldsymbol{F}$;
4. **Step 4:** Calculate the Jacobian matrix $\boldsymbol{J}$ and solve $\boldsymbol{J}\Delta \boldsymbol{U} = -\Delta \boldsymbol{F}$;
5. **Step 5:** Update: $\boldsymbol{U} \leftarrow \boldsymbol{U} + \Delta \boldsymbol{U}$;
6. **Step 6:** Output: $\boldsymbol{i} = g_2 \cdot \boldsymbol{U}[:, N]$;

[15] and solve the equation using an algorithm similar to Algorithm 1 and Algorithm 2, replacing conventional iterative methods.

$$g_2 \boldsymbol{D}_1 \left(\frac{1}{\boldsymbol{G}} \circ \boldsymbol{UD}\right) + \frac{g_2}{g_1} \boldsymbol{UD} + \boldsymbol{D}_1 \boldsymbol{U} - \boldsymbol{D}_1 \boldsymbol{V}_{ideal} = 0 \quad (15)$$

where $\boldsymbol{D}_1 = \begin{bmatrix} 2 & -1 & 0 & \cdots & 0 & 0 \\ -1 & 2 & -1 & \cdots & 0 & 0 \\ 0 & -1 & 2 & \cdots & 0 & 0 \\ \vdots & \vdots & \vdots & \ddots & \vdots & \vdots \\ 0 & 0 & 0 & \cdots & 2 & -1 \\ 0 & 0 & 0 & \cdots & -1 & 1 \end{bmatrix}$. For Eq. (15), its Jacobian matrix is

$$\boldsymbol{J}_{mvm} = (\boldsymbol{I}_N \otimes \boldsymbol{D}_1) diag\left(\frac{g_2}{\boldsymbol{G}}\right)(\boldsymbol{D} \otimes \boldsymbol{I}_N) \\ + \frac{g_1}{g_2}(\boldsymbol{D} \otimes \boldsymbol{I}_N) + (\boldsymbol{I}_N \otimes \boldsymbol{D}_1) \quad (16)$$

Consequently, the solution of Eq. (15) can be expressed as

$$\boldsymbol{U} = reshape[-\boldsymbol{J}_{mvm}^{-1}(vec(\boldsymbol{U})) \cdot \boldsymbol{F}(vec(\boldsymbol{U}))] \quad (17)$$

Compared with the node equation in [15], Eq. (15) contains only the tridiagonal matrices $\boldsymbol{D}$ and $\boldsymbol{D}_1$, without their inversion, and thus its Jacobian matrix exhibits greater sparsity. A summary of the MVM algorithm is shown in Algorithm 3.

### III. ACCURACY AND TIME COMPLEXITY ANALYSIS

*A. Accuracy of Algorithms*

We validated the accuracy of our algorithms by comparing their outputs with those obtained from SPICE simulations. The results validate our assumption that the errors stemming from the ideal OA approximations—namely, the "virtual short" and "virtual open"—are negligible. To ensure the convergence, the test matrices were randomly generated to be positive definite [4] and were subsequently mapped to conductance values ranging from 10 μS to 100 μS. Under these conditions, this comparison was performed across various matrix sizes, assuming a uniform interconnect resistance of 1 Ω for all rows and columns. As shown in Fig. 3, the strong consistency between the results of our proposed algorithm and those from SPICE simulations demonstrates the effectiveness of the INV circuit algorithm. Each of the four subplots presents 10 data sets for matrices of the corresponding sizes. As SPICE is highly computationally expensive, we present results only for matrices with a maximum size of 64×64 for comparison with the algorithm. Although the relative error (RE) increases with larger matrix size, it remains within an acceptable range (below $10^{-4}$).

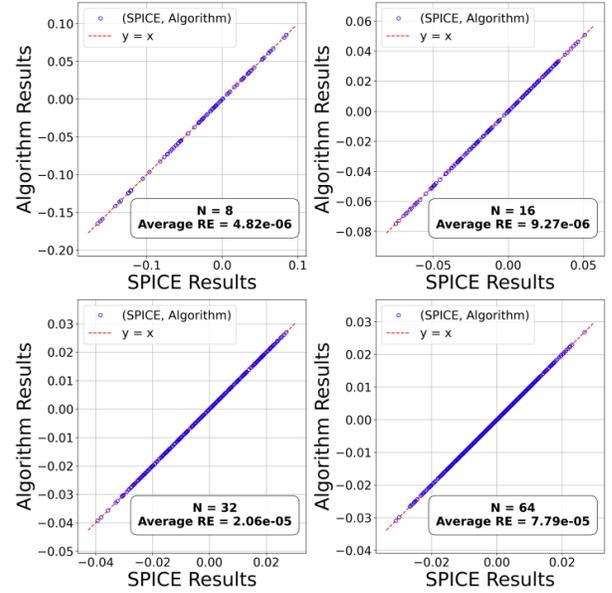

Fig. 3. Algorithm results as a function of SPICE results for INV circuit. The inset of each subplot shows the matrix size and the average relative error (RE) of 10 datasets calculated using the Euclidean norm. The red dashed line indicates the line where the horizontal and vertical coordinates are equal.

We comprehensively evaluate the relative error between the INV algorithm and the SPICE simulation under different matrix sizes and interconnect resistance values corresponding to a range of technology nodes, as shown in Fig. 4(a). In addition to the 1 Ω interconnect resistance, interconnect resistances of 1.55 Ω, 2.97 Ω and 4.53 Ω are also considered, corresponding to the 32 nm, 22 nm and 16 nm technology nodes, respectively, based on a $4F^2$ RRAM cross-point array structure, where $F$ denotes the feature size of the RRAM device [2][19]. As device dimensions shrink, interconnect resistance tends to increase, leading to a corresponding rise in relative error. Nevertheless, as shown in Fig. 4(a), when the matrix size does not exceed 64×64, the relative error remains below $10^{-3}$. The accuracy of the EGV and MVM circuit algorithm is comparable to that of the INV circuit. As shown in Fig. 4(b) and Fig. 4(c), its relative error is lower than $1\times 10^{-3}$, which strongly demonstrates the reliability of our algorithms.

*B. Time Complexity Analysis*

For the algorithms we proposed, the critical step determining their time complexity is the solution of $\boldsymbol{J}\Delta\boldsymbol{\theta} =$ $-\Delta F$, corresponding to Eq. (8) (Step 4 in Algorithm 1). When $N$ is sufficiently large (*e.g.*, $N \geq 128$), this step typically accounts for over 90% of the algorithm's total execution time. Therefore, we estimated the overall time complexity of the algorithms by analyzing this particular step.

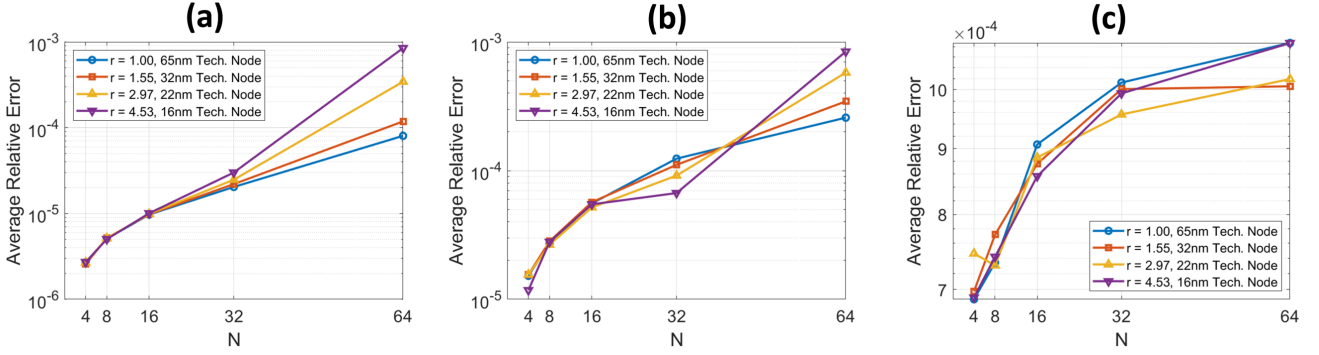

Fig. 4. Average relative error between the algorithms and SPICE with respect to increasing matrix sizes ($N$ = 4, 8, 16, 32, 64) and interconnect resistance values under different technology nodes. (a) Accuracy of the algorithm for the INV circuit. (b) Accuracy of the algorithm for the EGV circuit. (c) Accuracy of the algorithm for the MVM circuit.

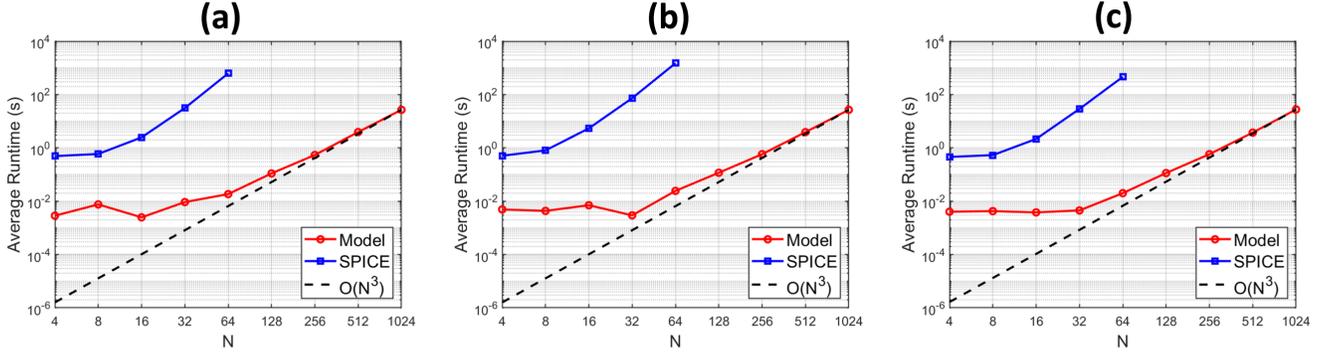

Fig. 5. Runtime comparison between the algorithms, SPICE and $O(N^3)$ reference line with respect to increasing matrix sizes. Owing to the high time complexity of SPICE, the AMC circuits simulated here are limited to matrix dimensions of at most $N$=64. (a) Runtime of the algorithm for the INV circuit. (b) Runtime of the algorithm for the EGV circuit. (c) Runtime of the algorithm for the MVM circuit.

This step fundamentally involves a matrix inversion problem, where the unknown $\Delta \boldsymbol{\theta}$ is a $N^2 \times 1$ vector, and $\boldsymbol{J}$ is an $N^2 \times N^2$ matrix. At first glance, this might suggest a high computational cost. However, in both Algorithm 1 and Algorithm 2, the vast majority of entries in $\boldsymbol{J}$ are zero, indicating a high degree of sparsity, as shown in Fig. 6. This sparsity enables highly efficient storage and computation [23].

For Algorithm 1, the expression for $\boldsymbol{J}$ is given in Eq. (9). Recalling the definitions of each term from earlier sections, it can be observed that the term $(\boldsymbol{I}_N \otimes \boldsymbol{D})diag\left(\frac{g_1}{G}\right)(\boldsymbol{M}_1\boldsymbol{D} \otimes \boldsymbol{I}_N)$ contributes the majority of non-zero entries in $\boldsymbol{J}$, specifically approximately $9N^2$ entries, i.e., $nnz(\boldsymbol{J}_{inv}) \approx 9N^2$, where $nnz$ denotes the number of non-zero entries. For Algorithm 2 and Algorithm 3, the expression for $\boldsymbol{J}$ is given in Eq. (12) and Eq. (16), respectively. The terms $(\boldsymbol{I}_N \otimes \boldsymbol{D})diag\left(\frac{g_1}{G}\right)(\boldsymbol{D} \otimes \boldsymbol{I}_N)$ and $(\boldsymbol{I}_N \otimes \boldsymbol{D}_1)diag\left(\frac{g_2}{G}\right)(\boldsymbol{D} \otimes \boldsymbol{I}_N)$ similarly contribute the majority of non-zero entries, again approximately $9N^2$ entries, i.e., $nnz(\boldsymbol{J}_{egv}) \approx 9N^2$ and $nnz(\boldsymbol{J}_{mvm}) \approx 9N^2$.

When $N > 32$, the sparsity of these matrices would be more than 99%. Consequently, the time complexity for solving the system of equations (Eq. (8), Eq. (13) and Eq. (17)) using LU factorization is approximately $O(f \cdot nnz)$ [21][22], which simplifies to $O(N^{2+\gamma})$, where $f$ and $\gamma$ are fill-in factors depending on the sparsity and matrix structure. In practical validation, due to large constant factors and the fill-in process during the factorization, the actual complexity is closer to $O(N^3)$, as will be revealed by our fitting results later. For the remaining steps, the time complexity is simply proportional to the number of non-zero elements in matrix $\boldsymbol{J}$, approximately $O(N^2)$.

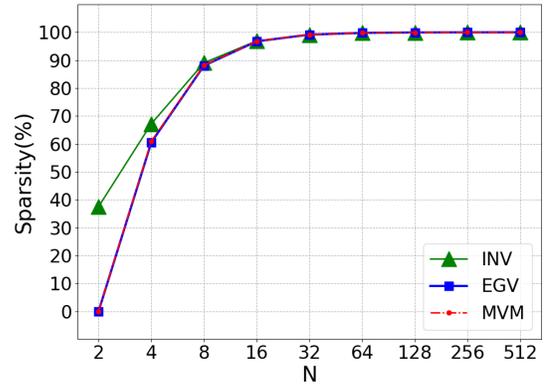

Fig. 6. Sparsity of the Jacobi matrices for three circuits (INV, EGV, MVM) as a function of matrix size $N$. Sparsity is defined as the ratio of zero elements to the total number of elements.

To evaluate the speedup of the proposed algorithms over conventional SPICE simulators, we conducted a series of simulations using Python3.11 and LTspice(R) XVII simulator on Intel(R) Core (TM) i7-11800H CPU @2.30GHz and 16GB RAM. Fig. 5 present the runtime curves for algorithms of INV, EGV and MVM circuits. The results demonstrate that these algorithms can estimate the impact of interconnect resistance for a 1024×1024 matrix within a few seconds on the given hardware platform. For a 64×64 matrix, their low time complexity enables a speedup of four orders of magnitude over SPICE. The curves also reveal that for small matrix dimensions $N$, runtime does not increase monotonically with $N$. This occurs because, at small $N$, nonzero entries from additional terms in the Jacobian become significant and cannot be neglected. Consequently, the growth in the number of nonzero elements is slower than $O(N^2)$, reducing the average nonzero entries per row and column as $N$ increases and thereby mitigating the computational cost of larger matrices.

Although sparse optimization can also reduce the computational cost of SPICE simulations [26], our algorithm retains a clear advantage for three main reasons: (i) SPICE processes matrices more than twice the size of those in our method; (ii) SPICE relies on Newton–Raphson iteration to handle nonlinear elements, which adds time overhead and often suffers from poor convergence with high-gain components; and (iii) nodes such as power supplies and common ground appear as dense rows and columns in sparse matrices, leading to substantial fill-in [25].

## IV. Compensating the Impact of Interconnect Resistance with the Proposed Algorithms

Thanks to their extremely low time complexity, the proposed algorithms can be applied to studies of interconnect resistance, including compensation strategies. For open-loop MVM circuits, several methods have been reported to address interconnect effects [14][24]. Building on our fast algorithms, we develop a bias-based strategy for closed-loop INV and EGV circuits to correct errors caused by interconnect resistance. This approach effectively reduces such errors, and we further determine the optimal bias that minimizes them while analyzing its dependence on matrix size and interconnect resistance.

For the INV circuit with interconnect resistances, the incorporation of interconnect resistances can be viewed equivalently that reduces the overall equivalent conductance $A$ in Eq. (1). Due to the reduction in $A$, the input current $b$ must be correspondingly reduced to ensure the correct output voltage $x$, implying a negative bias in the input current. We introduce a bias at the input port to modify the input current. The corresponding bias is proportional to $b$, denoted as $\delta b$. Using the proposed INV circuit algorithm, we evaluate how the relative error between the circuit's output voltage and the ideal solution varies with different levels of input current bias. The relative error corresponding to an output voltage of $x$ is defined as

$$\text{RE}_{inv}(x) = \frac{\|x - x_{ideal,inv}\|_2}{\|x_{ideal,inv}\|_2} \quad (18)$$

**Algorithm 4:** Search for Optimal Bias $\delta b$ to Minimize RE

1  Initialize Step as 0.02, offset as 0;
2  **for** $iter \leftarrow 1$ **to** 3 **do**
3      Divide Step by 10;
4      Initialize min(RE) $\leftarrow$ 9999, $i_{\min} \leftarrow 0$;
5      **for** $i \leftarrow 0$ **to** 19 **do**
6          $\frac{\delta b}{b} \leftarrow$ offset $+ (i - 15) \cdot$ Step;
7          Initialize RE $\leftarrow 0$;
8          **for** $j \leftarrow 1$ **to** 50 **do**
9              Set scaled current $b' \leftarrow b \cdot (1 + \frac{\delta b}{b})$;
10             $x_{\text{ideal}} \leftarrow A^{-1}b$;
11             $x \leftarrow$ Algorithm 1$(N, D, b', A, g_1, g_2, M_1, M_2)$;
12             RE $\leftarrow$ RE $+ \frac{\|x - x_{\text{ideal}}\|_2}{\|x_{\text{ideal}}\|_2}$;
13         **end**
14         Average the error: RE $\leftarrow$ RE/50;
15         **if** RE $<$ min(RE) **then**
16             min(RE) $\leftarrow$ RE, $i_{\min} \leftarrow i$;
17         **end**
18     **end**
19     offset $\leftarrow$ offset $+ (i_{\min} - 15) \cdot$ Step;
20 **end**
21 Set optimal bias $\delta b(\min(\text{RE})) \leftarrow b \cdot$ offset;
22 **return** $\delta b(\min(\text{RE})), \min(\text{RE})$;

where $x_{ideal,inv}$ is the analytical solution of Eq. (1). The absolute error is defined as the Euclidean norm of the difference between the output voltage and the analytical solution. Based on this evaluation, we identify the optimal input current bias that minimizes the relative error using Algorithm 4. We first update $\delta b$ using a coarse-grained approach with a large step size over a wide range. For each $\delta b$, we compute the average relative error across multiple input cases and identify the $\delta b$ corresponding to the minimum relative error. Then, around this $\delta b$, we narrow the search range and apply a fine-grained update using a smaller step size to find a new $\delta b$ that yields the minimum relative error. This process is repeated three times, each time narrowing the range and reducing the step size. The $\delta b$ corresponding to the final minimum relative error is taken as the optimal input current bias.

As shown in Fig. 7, the optimal input current bias is a small negative value, indicating that a slight reduction in input current results in the minimum relative error. The relative reduction $\Delta\text{RE}_{inv}$ is defined as

$$\Delta\text{RE}_{inv} = \frac{\text{RE}_{inv}(x_{\delta b=0}) - min(\text{RE}_{inv}(x))}{\text{RE}_{inv}(x_{\delta b=0})} \quad (19)$$

where $x_{\delta b=0}$ represents the output voltage without input current bias. The relative error at the optimal point is reduced by more than 50% compared to the relative error without compensation, demonstrating the effectiveness of the proposed compensation strategy in mitigating the impact of interconnect resistance. It should be noted that, based on our tests, this compensation strategy applies exclusively to the conductance matrix that is both diagonally dominant and symmetric. The variation of the optimal input current bias with matrix size and interconnect resistance in the INV circuit is shown in Fig. 8.

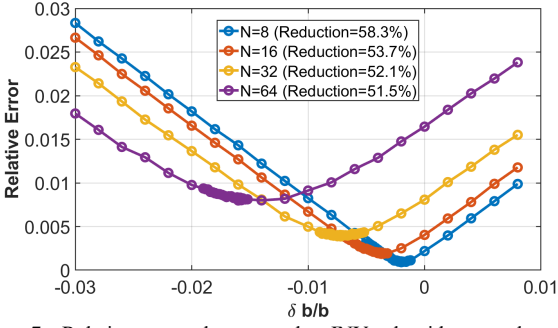

Fig. 7. Relative error between the INV algorithm results and analytical solution as a function of input current bias, evaluated under different matrix sizes with interconnect resistance set to 4.53 Ω. The figure highlights the optimal bias point that minimizes the relative error and presents the error reduction ratio relative to the point without bias. $b$ represents any element in the vector $\boldsymbol{b}$.

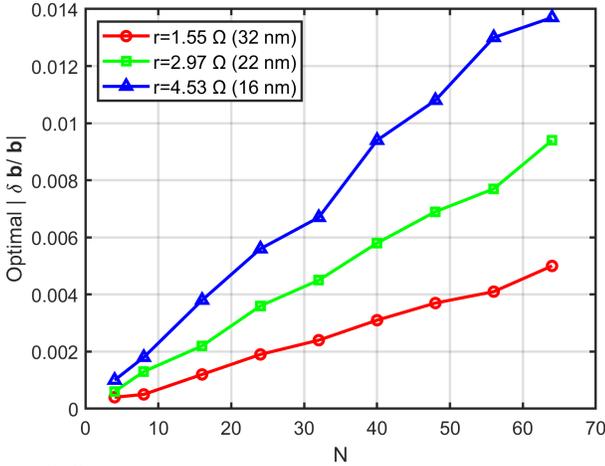

Fig. 8. Optimal current bias ratio of the INV circuit as a function of matrix size, shown for three typical interconnect resistance values corresponding to different technology nodes. $b$ represents any element in the vector $\boldsymbol{b}$.

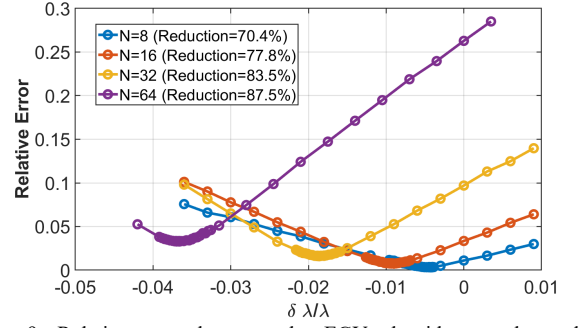

Fig. 9. Relative error between the EGV algorithm results and analytical solution as a function of eigenvalue bias, evaluated under different matrix sizes with interconnect resistance set to 4.53 Ω. The figure highlights the optimal bias point that minimizes the relative error and presents the error reduction ratio relative to the point without bias.

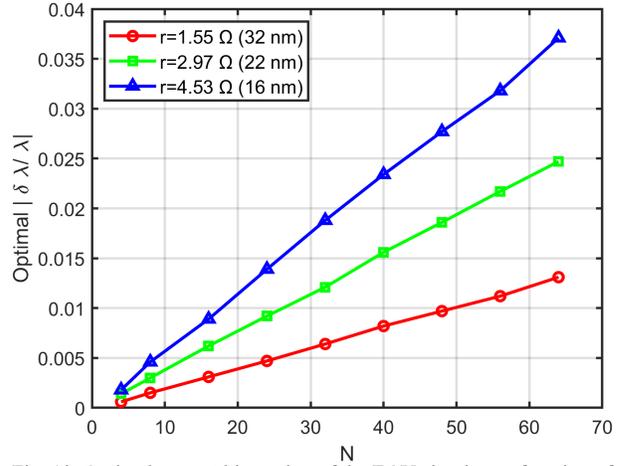

Fig. 10. Optimal current bias ration of the EGV circuit as a function of matrix size, shown for three typical interconnect resistance values corresponding to different technology nodes.

The optimal input current bias ratio increases with increasing matrix size and interconnect resistance, as larger matrices and higher interconnect resistance lead to greater errors, which require larger bias for effective compensation.

Different from the INV circuit, the EGV circuit has no explicit input and the known value is the eigenvalue $\lambda$ in Eq. (10). In this equation, the reduction in the equivalent conductance $\boldsymbol{A}$ caused by interconnect resistances necessitates a corresponding decrease in the eigenvalue $\lambda$ for compensation, namely, an eigenvalue bias $\delta\lambda$ is introduced, as shown in Fig. 9. At the optimal point, the proposed compensation reduces the relative error by over 70% compared with the uncompensated case. Fig. 10 shows how the relative error between the EGV circuit algorithm's result and the analytical solution varies with different levels of normalized eigenvalue bias. The relative error corresponding to an output voltage of $\boldsymbol{x}$ is defined as

$$\mathrm{RE}_{egv}(\boldsymbol{x}) = \left\| \frac{\boldsymbol{x}}{\|\boldsymbol{x}\|_2} - \frac{\boldsymbol{x}_{ideal,egv}}{\|\boldsymbol{x}_{ideal,egv}\|_2} \right\|_2 \quad (20)$$

where $\boldsymbol{x}_{ideal,egv}$ is the analytical solution of Eq. (10). The relative reduction $\Delta\mathrm{RE}_{egv}$ is defined as

$$\Delta\mathrm{RE}_{egv} = \frac{\mathrm{RE}_{egv}(\boldsymbol{x}_{\delta\lambda=0}) - min\left(\mathrm{RE}_{egv}(\boldsymbol{x})\right)}{\mathrm{RE}_{egv}(\boldsymbol{x}_{\delta\lambda=0})} \quad (21)$$

where $\boldsymbol{x}_{\delta\lambda=0}$ represents the output voltage without eigenvalue bias. Fig. 11 shows how the optimal eigenvalue bias changes with matrix size and interconnect resistance. Compared to the INV circuit, the EGV circuit shows a greater reduction in the relative error at the optimal point, while the optimal eigenvalue bias also increases with increasing matrix size and interconnect resistance. Similar to the INV circuit, a significant error reduction after compensation can only be achieved when the matrix is diagonally dominant and symmetric. We also apply a similar algorithm to the MVM circuit to determine an optimal input voltage bias for compensating errors caused by interconnect resistances. The dependence of optimal input voltage bias on matrix size and interconnect resistance is shown in Fig. 11. Therefore, this bias-based strategy is suitable for both closed-loop and open-loop AMC circuits.

## V. CONCLUSION

In this work, we derive the solutions for closed-loop INV

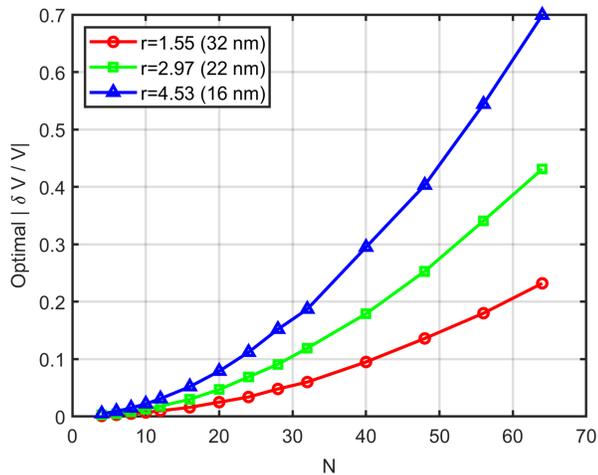

Fig. 11. Optimal current bias ration of the MVM circuit as a function of matrix size, shown for three typical interconnect resistance values corresponding to different technology nodes.

and EGV circuits with interconnect resistances and propose fast solving algorithms to accelerate computation. The algorithms for both circuits demonstrate efficiency in terms of time complexity, while achieving accuracy comparable to that of SPICE simulators. Due to their low time complexity, the algorithms outperform SPICE by several orders of magnitude. Moreover, a similar algorithm is also applicable to MVM circuits with interconnect resistances, where it has likewise proven effective and capable of significantly accelerating computation. Leveraging the high-speed algorithms, we develop a bias-based strategy to compensate for errors caused by interconnect resistance. It is demonstrated that this strategy effectively reduces the error at the optimal bias point and that the optimal bias increases with both matrix size and interconnect resistance.